\pdfoutput=1
\documentclass[conference]{IEEEtran}
\IEEEoverridecommandlockouts

\usepackage{cite}
\usepackage{amsmath,amssymb,amsfonts}
\usepackage{algorithmic}
\usepackage{graphicx}
\usepackage{textcomp}
\usepackage{xcolor}
\def\BibTeX{{\rm B\kern-.05em{\sc i\kern-.025em b}\kern-.08em
    T\kern-.1667em\lower.7ex\hbox{E}\kern-.125emX}}
\usepackage{tabularx}
\newcolumntype{Y}{>{\centering\arraybackslash}X}
\usepackage{multirow}
\usepackage{footnote}
\usepackage{placeins}
\usepackage{hyperref}
\usepackage{amssymb}
\usepackage{amsmath}
\usepackage{amsfonts}
\hypersetup{
  colorlinks = true,
  linkcolor  = blue,
  urlcolor   = blue,
  citecolor  = black,
}
\usepackage{url}
\usepackage{cleveref}
\usepackage{thmbox}
\usepackage{mdframed}
\usepackage{dblfloatfix}
\usepackage[acronym]{glossaries}

\newacronym{aiaa}{AIAA}{American Institute of Aeronautics and Astronautics}
\newacronym{dlr}{DLR}{Deutsches Zentrum für Luft- und Raumfahrt}
\newacronym{nasa}{NASA}{National Aeronautics and Space Administration}
\newacronym{orkg}{ORKG}{Open Research Knowledge Graph}
\newacronym{se2a}{SE²A}{Sustainable and Energy-Efficient Aviation}
\newacronym{tu}{TU}{Technische Universität}

\newcommand*{\fullref}[1]{\hyperref[{#1}]{\cref*{#1} \nameref*{#1}}}
\newcommand*{\Fullref}[1]{\hyperref[{#1}]{\Cref*{#1} \nameref*{#1}}}
\newcommand*{\secref}[1]{\hyperref[{#1}]{\autoref*{#1}}}
\newcommand*{\Secref}[1]{\hyperref[{#1}]{\Cref*{#1}}}

\let\oldFootnote\footnote
\newcommand\nextToken\relax

\renewcommand\footnote[1]{%
    \oldFootnote{#1}\futurelet\nextToken\isFootnote}

\newcommand\isFootnote{%
    \ifx\footnote\nextToken\textsuperscript{,}\fi}

\usepackage{booktabs}
\usepackage{xcolor}
\def\BibTeX{{\rm B\kern-.05em{\sc i\kern-.025em b}\kern-.08em
    T\kern-.1667em\lower.7ex\hbox{E}\kern-.125emX}}

\newcounter{openboxwithtitle}[section] 
\newenvironment{openboxwithtitle}[1]{
  \refstepcounter{openboxwithtitle} 
  \vspace{1.75ex}
  \thmbox[L]{\textbf{#1}}
  \hspace*{-1.5em}\slshape\ignorespaces%
}{
  \endthmbox\vspace*{.75ex}
}

\usepackage[all]{nowidow}

\begin{document}

\title{Aerospace.Wikibase: Towards a Knowledge Infrastructure for Aerospace Engineering
\thanks{This work was funded by the DFG SE2A Excellence Cluster.}
}

\author{
    \IEEEauthorblockN{
    Tim Wittenborg\IEEEauthorrefmark{1},
    Ildar Baimuratov\IEEEauthorrefmark{1},
    Jamal Eldemashki
    \IEEEauthorblockA{\IEEEauthorrefmark{1}L3S Research Center, Leibniz University Hanover, Hanover, Germany
    \\\{tim.wittenborg, ildar.baimuratov\}@l3s.uni-hannover.de}
}
}

\maketitle

\begin{abstract}
While Aerospace engineering can benefit greatly from collaborative knowledge management, its infrastructure is still fragmented.
Bridging this divide is essential to reduce the current practice of redundant work and to address the challenges posed by the rapidly growing volume of aviation data.
This study presents an accessible platform, built on Wikibase, to enable collaborative sharing and curation of aerospace engineering knowledge, initially populated with data from a recent systematic literature review.
As a solid foundation, the Aerospace.Wikibase provides over 700 terms related to processes, software and data, openly available for future extension.
Linking project-specific concepts to persistent, independent infrastructure enables aerospace engineers to collaborate on universal knowledge without risking the appropriation of project information, thereby promoting sustainable solutions to modern challenges while acknowledging the limitations of the industry.
\end{abstract}


\section{Introduction}
The aerospace industry integrates complex systems and cutting-edge research to drive progress in design, manufacturing, and operational efficiency.
However, the potential of flexible data and information reuse has yet to be fully realized in the aerospace domain.
Studies highlight that, despite the aerospace industry's longstanding tradition of applying knowledge-based methodologies, significant gaps in interoperability, standardization, and adoption still persist\cite{harvey_knowledge_2005,sanya_challenges_2011}.
Limitations in cross-disciplinary collaboration and knowledge sharing, the absence of unified data standards and insufficient compliance with the FAIR principles\cite{wilkinson_fair_2016}, \textbf{F}indability, \textbf{A}ccessibility, \textbf{I}nteroperability, and \textbf{R}eusability, hinder the realization of knowledge-based engineering’s potential.
Moreover, the divide between domain-specific engineers and knowledge engineers emphasizes the need for collaborative frameworks that bridge practical and theoretical knowledge applications\cite{haney_data_2016,procko_leveraging_2022}.
Implementing these standards can promote innovation, optimize workflows, and advance sustainability objectives.

Our goal is to identify a more sustainable approach to crowdsourcing aerospace engineering knowledge beyond project timelines and institutional boundaries.
Our resulting research question is:
\newpage
\begin{openboxwithtitle}{Research Question}
~How can aerospace engineers share their knowledge on openly accessible and sustainably lasting infrastructure?
\end{openboxwithtitle}

Our contribution includes:
\textbf{1)} An overview of prior approaches to aerospace engineering knowledge exchange.
\textbf{2)} An explicit data model derived from a recent systematic literature review on knowledge-based aerospace engineering.
\textbf{3)} The implementation of the Aerospace.Wikibase as an open and persistent knowledge infrastructure.

The remaining paper is structured as follows:
\Secref{sec:background} establishes the state of the art in aerospace knowledge infrastructures, and \secref{sec:approach} illustrates our approach to improve upon it.
\Secref{sec:results} provides details on the developed Wikibase instance.
These findings are discussed in \secref{sec:discussion}, whereas \secref{sec:conclusion} concludes the paper.

\section{Background and Related Work\label{sec:background}}
Knowledge Infrastructures (KI) are defined by Edwards\cite{10.5555/1805940} as \textit{``...robust networks of people, artifacts, and institutions that generate, share, and maintain specific knowledge about the human and natural worlds''}.
The Wikipedia Community is a notable example, where the Wikimedia Foundation (institution) provides a digital infrastructure (artifact) to bring together ``\textit{a diversity of actors, organizations and perspectives from, for instance, academia, industry, business and general public}''\cite{karasti2016knowledge} (people).
A wide range of institutions in aerospace engineering is easily identified, ranging from organizations such as \gls*{nasa} and Siemens to projects like AGILE\footnote{\url{https://www.agile4.eu/}}, \gls*{se2a}\footnote{\url{https://www.tu-braunschweig.de/en/se2a}}, and SynTrac\footnote{\url{https://www.trr-syntrac.com/}}.
Artifacts, particularly openly accessible ones, are much harder to identify, since ``\textit{the aerospace and defense industries are perhaps the most governed of all sectors, being regulated by a plethora of agencies, both national and regional}'', according to Harvey and Holdsworth~\cite{harvey_knowledge_2005}.
While various knowledge management systems and ontologies have been proposed for the aerospace domain \cite{dadzie_applying_2009,curran_knomad_2010,sanya_ontology_2014,procko_leveraging_2022}, there are, to our knowledge, no open-access collaborative aerospace knowledge bases.
The nearest equivalent may be the glossary of aerospace engineering\footnote{\url{https://en.wikipedia.org/wiki/Glossary_of_aerospace_engineering}} on Wikipedia.
All other open and collaborative knowledge bases seem to have been discontinued after their individual project time\cite{karasti_infrastructure_2010}.

This sentiment is shared in the systematic literature review on knowledge-based aerospace engineering by \cite{wittenborg2025knowledge}, concluding that knowledge is usually not formalized, but distributed on the human resources level, in reports, meetings, and direct communication, and closed to the outside.
To address this, the authors extracted a dataset of terms related to aerospace engineering processes, software and data, and ingested it into the \gls*{orkg}\cite{auer2020improving}, a platform that enables structured, machine-readable representation of research contributions.
This makes it easier to compare, search, and interlink scientific knowledge across disciplines.
Although this knowledge infrastructure is openly available for collaboration, the \gls*{orkg}’s research contribution–centric model seems too restrictive for engineers, who need a system to standardize domain terms and their interconnections.

As such, an advance to a dedicated Wikibase.Cloud instance is recommended.
Wikibase\footnote{\url{https://wikiba.se/}} is free and open-source software for storing and managing structured, linked, and collaborative data, similar to Wikidata. 
Like the \gls*{orkg}, Wikibase represents content in a structured, machine-readable way, but it is not limited to a specific data model and is suitable for sharing aerospace engineering knowledge that is not formulated as research contributions.
However, Wikibase is only software and to use it for sharing aerospace engineering knowledge, it must be properly configured and deployed.

\section{Approach\label{sec:approach}}
We aim to close this gap in the state of the art of knowledge infrastructure in the aerospace domain by developing the \href{https://aerospace.wikibase.cloud}{Aerospace.Wikibase.Cloud} platform.
This primarily involves two major steps: (1) data acquisition and modeling, (2) and implementation and population of the Wikibase instance.

\subsection{Data Acquisition and Modeling\label{sec:data}}
We reused the data from the systematic literature review\cite{wittenborg2025knowledge}, which identified over 700 domain-specific aerospace engineering terms and their interconnections.
A semantic model was designed to structure this data for the Aerospace.Wikibase under development.
The model includes several classes derived from the following research question:
\textit{``Which Aerospace Engineering \textbf{Process} is completed by which \textbf{Software} using which \textbf{Data}, in which \textbf{Format}, and with which \textbf{Schema}?''}
For \texttt{Data format specification} and \texttt{Data item}, we import the corresponding classes from the Information Artifact Ontology\cite{iao}.
The \texttt{Data model} class is linked to a Wikidata entity\footnote{\url{https://www.wikidata.org/wiki/Q1172480}}.
The \texttt{Process} class is imported from BFO, and the \texttt{Software} class is imported from the Software Ontology\cite{swo}.
In addition to these classes and following the ORKG data model, we introduced the \texttt{Contribution} class, which represents academic publications presenting research contributions in the aerospace domain.

The model includes several relations. Most of these define connections between instances of the classes, such as \texttt{has process}, \texttt{has software}, \texttt{has data item}, \texttt{has data format specification}, and \texttt{has data model}. In addition, the model includes the relation \texttt{mentions}, which connects a \texttt{Contribution} to the domain entities referenced in it. Another relation we introduce is \texttt{has part}, which connects, for example, processes with their subprocesses. Finally, we use the relation \texttt{instance of} to establish instance–class relationships between domain entities (e.g., a 3D model is an instance of a model).

\begin{figure}
    \centering
    \includegraphics[width=\linewidth]{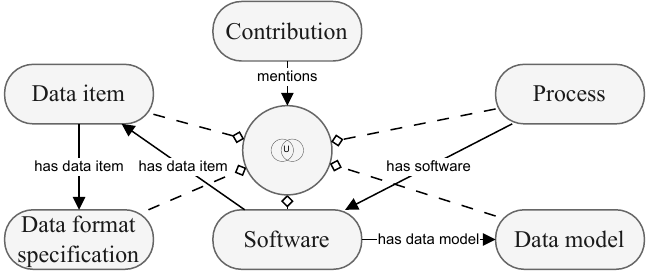}
    \caption{Data model}
    \label{fig:onto}
\end{figure}

The model also employs several annotation properties. The first, \texttt{alias}, is used to store alternative or synonymous names for the same entity. The second, \texttt{source}, denotes the origin of the information, such as literature review, survey, or interview. Finally, entities are aligned with Wikidata URIs to enhance the FAIR compliance of the Wikibase content. \autoref{fig:onto} depicts the designed data model, which is available as an OWL file.

\subsection{Wikibase Implementation and Population}
The aerospace knowledge base was built on a dedicated Wikibase instance using a structured workflow comprising system deployment and automated data ingestion.
In addition to the generic features of Wikibase, we developed a tailored approach to ensure that the data was imported consistently, preserving the class hierarchies while leveraging Wikibase flexibility and user-interface.

\subsubsection{Mapping to Wikibase}
The OWL constructs of the data model were mapped to the Wikibase data model as follows:
\begin{itemize}
    \item \textbf{Classes} $\rightarrow$ Wikibase items, with \texttt{rdfs:subClassOf} mapped to the \texttt{subclass of} property.
    \item \textbf{Individuals} $\rightarrow$ items linked via \texttt{instance of}, where each \texttt{rdf:type} triple was converted accordingly.
    \item \textbf{Object properties} $\rightarrow$ Wikibase properties with appropriate datatypes.
    \item \textbf{Labels, descriptions, and aliases} $\rightarrow$ Wikibase fields for searchability.
    \item \textbf{External identifiers and provenance} $\rightarrow$ Wikidata URIs, DOIs, and a dedicated \emph{ontology\_iri} property ensured traceability to the source ontology.
\end{itemize}
Wikibase exposes both items and properties as first-class entities with labels, descriptions, and documentation of use. 
Our mappings ensured that statements such as \texttt{instance of}, \texttt{subclass of}, \texttt{ontology IRI}, or \texttt{Wikidata URI} appear with clear semantics, making the imported ontology human-readable in the UI.

\subsubsection{Data ingestion}
The import pipeline was implemented in Python using \texttt{rdflib} for parsing and \texttt{WikibaseIntegrator} for ingestion, building on the recent outline by Stehr\cite{stehr_digitale_2025}.
Properties were created manually following Wikidata.org conventions (explicit datatypes, descriptive labels, usage notes), since they define the schema and must remain stable.
These properties were then referenced in a mapping dictionary that links ontology annotations to their Wikibase equivalents, ensuring consistent reuse.  

The ingestion proceeded in three stages.
First, all classes are imported, establishing the subclass hierarchy.
Then, all individuals are created and linked to their classes with \texttt{instance of}, enriched with additional metadata such as \texttt{ontology\_iri}, and assigned aliases stored as Wikibase aliases.
To conclude these steps, we made use of the Wikibase Query Service, which provides a SPARQL endpoint in the user interface, and ran SPARQL queries to verify the correct import of labels, hierarchy links, and identifiers.

\subsubsection{Data quality and error handling}
During ingestion, temporary failures occasionally occurred, such as \emph{database locked} or \emph{failed-save} errors caused by replication lag or API timeouts. To mitigate these issues, the bot was designed to be idempotent: each entity was first checked against a local cache and, when available, matched by external identifiers before creation.
This ensured that re-running the pipeline after a failure did not result in duplicate entities but resumed the ingestion from a consistent state.
After each batch, automated SPARQL queries were executed to verify data quality, checking for missing labels, duplicate items, and orphan classes without subclass or instance links.
These quality control loops allowed systematic validation of both content completeness and structural integrity.

\subsubsection{Incremental extensibility}
The system was designed to accommodate the integration of additional ontologies beyond the initial import.
By extending the mapping dictionary with new annotation or object properties, and by creating any missing Wikibase properties in the schema, new data sources can be incorporated without disrupting the existing knowledge base.
This incremental approach allows the knowledge graph to grow in a controlled and consistent manner, ensuring scalability and adaptability over time.
It also prevents schema fragmentation, since all ontologies must conform to the same property mapping strategy.

\section{Results\label{sec:results}}
The resulting Aerospace.Wikibase instance contains a total of 14,886 nodes forming 37,897 triples, of which 8,202 (55\%) nodes are IRIs and 6,660 (45\%) are literals.
The literals fall into four data types: 3,607 (54\%) of type \texttt{rdf:langString}, 1,271 of type \texttt{xsd:integer} (19\%), 954  (14\%) of type \texttt{xsd:string} and 828 (12\%) of type \texttt{xsd:dateTime}.
Among the triples, 15,513 (41\%) are about domain properties described in \autoref{sec:data}, 10,393 (27\%) relate to Wikibase ontology, including \texttt{rank}, \texttt{timestamp}, \texttt{statements}, \texttt{sitelinks} and \texttt{identifiers}, and 11,990 (32\%) have other properties such as \texttt{rdf:type}, \texttt{skos:altLabel}, \texttt{schema:dateModified}, \texttt{schema:version}, \texttt{rdfs:label} and \texttt{schema:description}.
The use of domain items and properties is summarized in \secref{tab:res}.
With close to 1000 pages, Aerospace.Wikibase falls within the top 13th percentile of Wikibase instances by page count.

\begin{table}[bh]
    \centering
    \caption{Domain entities in the Aerospace.Wikibase. Note that ``Edge'' counts the amount of edges, not edge types.}
    \begin{tabularx}{\linewidth}{l|l|Y|Y}
        \multicolumn{2}{c|}{\textbf{Type}} & \textbf{Amount} & \textbf{Total} \\
        \midrule
        \multirow{3}{*}{Node} & Instance & 637 & \multirow{3}{*}{710} \\
        \cline{2-2}
        & Class & 7 &  \\
        \cline{2-2}
        & Ambiguous & 66 & \\
        \hline
        \multirow{3}{*}{Edge} & Object property (internal) & 1,533 & \multirow{3}{*}{4,466} \\
        \cline{2-2}
        & Object property (external) & 1,231 &  \\
        \cline{2-2}
        & Annotation property & 1,533 &  \\
    \end{tabularx}
    \label{tab:res}
\end{table}

\section{Discussion\label{sec:discussion}}
The Wikibase now provides a more flexible environment for data access and curation.
The ingested data had a strict classification into instances and classes. While this maintains the ability to connect, for example, software instances to process instances, these might otherwise have been represented using classes and subclasses.
Within Wikibase, this strict modelling is replaced with an emerging multi-layered structure, allowing for \texttt{instance of} connections between items that were previously treated as instances themselves.
For example, a 3D model is an instance, or potentially a subclass, of data item and model.
Presumably, most of these will eventually converge in a multi-layered class- and meta-class-system.
Lifting these restrictions is one of the major benefits of the Wikibase environment, alongside the simple, familiar user interface.
Since it is connected to, yet not contained within Wikidata, engineers can now model and extend their domain knowledge without familiarizing themselves with Wikidata’s additional conventions. 

Providing the technical possibility may not, however, overcome the apparent inherent opposition in the engineering domain to share knowledge freely, particularly in aerospace engineering.
Years of previous work have identified this status of \textit{``the most governed of all sectors''}~\cite{harvey_knowledge_2005}, focused on acquiring and isolating knowledge to secure economic or even military advantage before the apparent opposition catches up.
Yet, the same works that identify this entrenched position also indicate that coupling beyond system, domain and institutional boundaries is becoming a necessity to overcome the increasingly complex, time-constrained challenges of the aerospace sector.
Projects such as AGILE, \gls*{se2a} or SynTrac show that these solutions are not undesired, but needed and potentially required to solve future problems.

Our Wikibase initiative builds not on the assumption that every engineer will contribute to a single platform, but rather on enabling them to collaboratively use common artifacts and knowledge bases.
Wikipedia has 2000 reads per edit\footnote{\url{https://en.wikipedia.org/wiki/Wikipedia:Statistics}}, being consumed much more than contributed to.
As long as domain engineers agree on a common language, their private collaboration can be linked over a publicly curated terminology. 
The Aerospace.Wikibase provides an example of how a sovereignly manageable yet fundamentally interconnected knowledge graph can serve as a central knowledge sharing infrastructure, fitted to domain specific requirements.
Ideally, such a Wikibase would be picked up by central institutions representing aerospace, such as \gls*{nasa}, where MediaWiki is already used for internal knowledge management, \gls*{aiaa}, \gls*{dlr}, \gls*{tu} Delft, or similar.
Independently, even a single engineer could populate Aerospace.Wikibase or their own Wikibase, or start contributing on Wikidata and the \gls*{orkg} right away.
The knowledge infrastructure, required by the previous authors, is available today.

\begin{openboxwithtitle}{Research Finding}
~Aerospace engineers can share their knowledge in the dedicated, openly accessible \href{https://aerospace.wikibase.cloud}{Aerospace.Wikibase.Cloud} instance.
This sustainably lasting infrastructure is independent of project funding and domain restrictions and provides a universal knowledge node to couple private systems.
\end{openboxwithtitle}

\section{Conclusion and Future Work\label{sec:conclusion}}
In this research, we addressed the lack of a collaborative, open, and sustainable knowledge infrastructure in the aerospace engineering domain by developing a dedicated Wikibase instance.
Using data from a recent systematic literature review that identified domain-specific terms and their interconnections, we modeled this information as an OWL ontology, reusing classes from the Basic Formal Ontology, the Information Artifact Ontology, the Software Ontology, and Wikidata.
We then implemented a workflow consisting of the system implementation and a programmatic interface for automated data ingestion.
This workflow ensures consistent data import, preserves hierarchical structures, and optimizes user interaction.
The resulting Wikibase contains 14,886 nodes forming 37,897 triples, of which 1,025 and 4,759 are related to domain-specific entities.
This shows that already a small amount of curative work can greatly advance the availability, and that over a hundred other platforms are already thriving further in the Wikibase ecosystem.
As such, we see two major angles for future work: first, fostering data scope and quality, and second, advancing aerospace engineering knowledge infrastructure.
The first can be achieved by integrating additional artifacts beyond our foundational dataset, while advancing the knowledge infrastructure requires people and institutions to use these artifacts to their advantage.
This future work is required to overcome the challenges in aerospace engineering knowledge management.

\section*{Acknowledgment}
\paragraph*{Use of AI tools declaration}
During the preparation of this work, the author(s) used \textbf{DeepL}, \textbf{Grammarly (Browser Plugin)}, \textbf{LanguageTool (Browser Plugin)} in order to: \textbf{translate text}, \textbf{grammar and spelling check}, \textbf{paraphrase and reword}, according to the CEUR GenAI Usage Taxonomy\footnote{\url{https://ceur-ws.org/GenAI/Taxonomy.html}}.
After using this tool/service, the authors reviewed and edited the content as needed and take full responsibility for the publication’s content.

\bibliographystyle{ieeetr}
\bibliography{main}

\end{document}